\documentstyle[12pt]{article}

\begin{document}
\hfill{LA-UR-93-832}

\vspace{7pt}
\begin{center}
{\large\sc{\bf A Candidate for Exact Continuum
 Dual Theory for Scalar QED$_3$.}}

\baselineskip=12pt
\vspace{35pt}

A. Kovner$^ *$ and P. Kurzepa $^{**}$\\
\vspace{10pt}
Theory Division, T-8,
Los Alamos National Laboratory,
MS B-285\\
Los Alamos, NM 87545\\
\vspace{5pt}
 and \\
\vspace{5pt}
B. Rosenstein$^{\dagger}$  \\
\vspace{10pt}
Institute of Physics,
Academia Sinica\\
Taipei 11529.,
Taiwan, R.O.C.\\
\vspace{35pt}
\end{center}
\begin{abstract}
We discuss a possible exact equivalence of the Abelian
Higgs model and a scalar theory of a magnetic vortex
field in 2+1 dimensions. The vortex model has a
current - current interaction and can be viewed as a
strong coupling limit of a massive vector theory.
The fixed point structure of the theory is discussed
and mapped into fixed points of the Higgs model.
\end{abstract}

\vspace{50pt}
\
\newline
*KOVNER@PION.LANL.GOV
\newline
**KURZEPA@PION.LANL.GOV
\newline
$\dagger$ BARUCH@PHYS.SINICA.EDU.TW
\vfill
\pagebreak

\section{Introduction}
Quantum electrodynamics in 2+1 dimensions has recently attracted
much attention as a model theory for different physical phenomena.
The theory with massless fermions is interesting as a model
of some aspects of QCD since it
exhibits spontaneous chiral symmetry breaking \cite{appel}.
When endowed with Chern - Simons term the model exhibits
the statistics transmutation phenomenon and is used as
an effective theory for description of the  Quantum Hall
Effect \cite{zhang}. Different variants of QED$_3$ have
been also used to describe the high temperature
superconductors \cite{prb}.

Apart from this the scalar QED$_3$ - the Abelian
Higgs model- has always been interesting as the Landau -
Ginzburg theory for ordinary superconductors. For example
the order of the Higgs - Coulomb phase transition for a
long time has been a matter of controversy
\cite{halperin},\cite{kleinert} and the model was studied
by a variety of numerical \cite{dasgupta},\cite{bartholomew}
and analytical \cite{halperin},\cite{kleinert},\cite{stone} methods.

Recently we have studied the Abelian Higgs model in 2+1
dimensions using continuum duality methods \cite{npb}.
The results of this analysis can be summarized as follows.
The Higgs - Coulomb phase transition can be understood as due
to condensation of magnetic vortices. The phase transition can
be described by a local order parameter - the vortex creation
operator $V(x)$.
It has a nonvanishing expectation value in the Coulomb phase
and zero VEV in the Higgs phase. The field $V(x)$ is a scalar
complex field. Its phase rotations are generated by the operator
of the magnetic flux $\Phi=\int d^2x B(x)$.
\begin{equation}
[V(x),\Phi]=\frac{2\pi}{e}V(x)
\end{equation}
This $U(1)$ transformation is a symmetry of the theory since
the magnetic flux is conserved. The corresponding conserved
current is the dual field strength $\tilde F_\mu=
\epsilon_{\mu\nu\lambda}\partial_\nu A_\lambda$ and its
conservation equation is just the homogeneous Maxwell
equation of the theory.
The nonvanishing VEV of $V(x)$ in the Coulomb vacuum
therefore signalls spontaneous breaking of the magnetic flux symmetry in
the Coulomb phase. The SSB of a global continuous symmetry is
followed by the appearance in the spectrum of a massless Goldstone
mode. Here it is the massless photon.
The electrically charged states of QED are topological solitons of the
vortex field $V(x)$ and asymptotically have the form
$V(x)\rightarrow_{r\rightarrow \infty}v\exp{i\theta}$ where
$\theta(x)$ is a polar angle.

 In the Higgs phase the flux symmetry is restored. The massless
excitation disappears from the spectrum and the operator $V(x)$
itself interpolates a massive particle - the magnetic vortex.

This simple picture suggests that the phase transition itself and
the low energy properties of the theory near the phase transition can
be described by a Landau theory written in terms
of the order parameter field $V(x)$.
\begin{equation}
L_{Landau}=\partial_\mu V\partial_\mu V^* -U(V^*V)
\label{landau}
\end{equation}
where the potential $U(V^*V)$ has a single minimum
at $V=0$ in the Higgs phase and degenerate minima $|V|=const$
in the Coulomb phase. Indeed it was shown in
\cite{npb} and \cite{prl} that the Lagrangian eq.(\ref{landau})
describes correctly the low energy properties of the theory
including the Aharonov - Bohm effect.
The standard universality argument then suggests that the
Coulomb - Higgs phase transition is in the same universality
class as the 3D XY model. The elementary "spin" of this XY model
is the vortex field $V$. Moreover the same argument suggests
that if the theory has a tricritical fixed point it must be
the theory of a free vortex field $V$.

The interesting possibility is that the notion of dual Lagrangian
may be more fundamental than just the effective low energy theory in
the sense of Landau. It may be possible to construct a
local Lagrangian in terms of the vortex field $V(x)$ only which
is exactly equivalent to the scalar $QED_3$ at all scales and not
just at low energies. Should such a Lagranian exist, it
would be a completely local description of $QED_3$ in terms of
local gauge invariant variables. In the following we will refer to
this model as Quantum Vortex Dynamics (QVD).

Some work has been done in this direction in the late seventies
and early eighties but the form of the exact dual Lagrangian is not
known. There are two approaches to the derivation of the QVD
Lagrangian. One is to attempt to perform an exact dual
transformation in the lattice regularized theory
 \cite{stone},\cite{kleinert}. In this way one can quite easily
dualize the Villain version of the fixed length Abelian Higgs model.
However this Villain model and the original theory are supposed
to be equivalent only near the infrared fixed point and therefore
this excercise does
not tell one much about the ultraviolet form of the dual theory.
Moreover it would probably be too ambitious to try to find an
exact equivalent on the lattice. The lattice dual theory could
in principle contain a host of irrelevant terms (some even nonlocal)
which would all disappear in the continuum limit and therefore
would be totally irrelevant for the continuum duality. Another
approach is therefore to perform a dual transformation
directly in the continuum \cite{bardakci}. However a formal dual
transformation in continuum has the same drawback. It generates
a complicated nonpolynomial
Lagrangian totally unappealing from both, aesthetic and computational
points of view. One expects again that most of its terms are irrelevant
and therefore
can be dropped with the only effect of renormalizing the
coefficients of the remaining terms. It is however not clear
which are the remaining relevant terms.

In this paper therefore we will take a different approach to
determining the exact dual theory. In Sec.2 we will
explain several conditions which the
dual theory must satisfy given that the dynamics of
the Abelian Higgs model
in the ultraviolet regime can be probed with the help
of perturbation theory.
We also describe the known fixed point structure of
the RG flow in the model.
In Sec.3 we present a model of the vortex field $V(x)$ which
satisfies all these conditions. We  discuss the fixed point
structure of this model and
show that it is consistent with the known (and expected) fixed
points of scalar QED$_3$ and the assumption of duality.
Sec. 4 is devoted to a short discussion.

\section{Necessary conditions and the fixed points.}
In this section we summarize the conditions that the
dual theory must satisfy
which can be inferred with a large measure of certainty
from the standard perturbative calculations in the framework
of QED. The Lagrangian of QED$_3$ is

\begin{equation}
L_{Higgs}=-\frac{1}{4 g^2}F_{\mu\nu}^2+|
(\partial_\mu+iA_\mu)\phi|^2+\mu^2\phi^*\phi-\kappa(\phi^*\phi)^2
\label{higgsl}
\end{equation}

1.The most obvious condition is that QVD must have a well
defined continuum limit or in the language of continuum theory
must be renormalizable.
This renormalizability does not have to be perturbative, in a
sense that some couplings of the dual theory may be not small but
rather take values in the vicinity of some finite fixed point.
In fact it is pretty obvious that this must be the case. If all
the couplings of the vortex model were small in the ultraviolet
(asymptotically free), the UV fixed point would be a free theory
of the vortex field. However the Abelian Higgs model in the UV region
is a theory of free photons and free charges rather than free vortices.

2. Consistent with the universality assumption in the low energy
region the QVD Lagrangian must reduce to the Landau theory
eq.(\ref{landau}). Formally this means that in addition to the usual
potential energy the interaction Lagrangian can have higher
derivative terms which are important in the UV (consistent
with the previous point) but are very small in the IR.

3. In the framework of the Abelian Higgs model the vortex
operator $V(x)$ is a local operator for any nonzero value of
the finite structure constant $e^2$.
The charged field $\phi(x)$ is on the other hand nonlocal in
any physical gauge.
In the limit $e^2\rightarrow 0$ however the charged field
decouples from the photon fields and becomes itself
gauge invariant and local. In the same limit
magnetic vortices (which are nonlocal configurations of $\phi$)
become nonlocal objects and the vortex operator $V(x)$ ceases
to be a local field. Another manifestation of the singular
nature of the limit $e^2\rightarrow 0$ is that the quantum
numbers of the vortex operator are changed. For any nonzero $e^2$
the vortex configuration that carries a nonzero vorticity
$\int d^2x \epsilon_{ij}\partial_i(\frac{\phi^*\stackrel
{\leftrightarrow}\partial_j\phi}{\phi^*\phi})$,  carries also
nonzero magnetic flux. For $e=0$ however the photon field
is decoupled and the vortices therefore do not carry magnetic
flux. In the
dual theory as discussed in \cite{prl} the charged field
creates topological
solitons of $V(x)$.
The dual theory should therefore have a well defined limit in
which these solitons become local and the field $V$ looses locality.

4. Another hint on the form of the QVD interaction can be
gathered from the spectrum of QED$_3$ in the Higgs phase.
Perturbatively the spectrum contains a massive vector
particle (massive photon) and a massive scalar Higgs. In terms
of the vortex model those should be evidently interpreted
as vortex - antivortex bound states in the disordered phase.
However a scalar theory with just potential interaction of the
type $(V^*V)^2$ can not have bound states. The interaction of
this type must be repulsive, since otherwise the theory
is unstable. Moreover in any theory of this type the
vortex-vortex and vortex-antivortex interactions are the
same which is not the case in QED$_3$. One simple
possibility to induce bound states is to have an interaction
of the vector-vector type
$(V^*\stackrel{\leftrightarrow}\partial_\mu V)^2$. In that case
one can have attraction in the vortex-antivortex channel and
repulsion in the vortex-vortex channel allowing for appearance
of bound states without destabilizing the theory.
Also, the massive photon and the Higgs particle in perturbation
theory are light whereas the vortices themselves are nonperturbatively
heavy. This means that the vector type interaction, if present
must be very strong for small electromagnetic coupling
$e^2$ to produce such a large binding energy.

5. Finally, another strong constraint comes from the known
ultraviolet behaviour of QED$_3$. In the ultraviolet the
theory is asymptotically free. One can therefore calculate the
UV behaviour of the vortex operator. This was done in
\cite{npb} with the following result
\begin{equation}
<TV(x)V^*(y)>\rightarrow_{|x-y|\rightarrow 0}|x-y|^a
 \exp\{\frac{\pi}{e^2|x-y|}\}
\label{uvas}
\end{equation}
This behaviour is somewhat unusual, since it is telling us that
in the UV region the field $V$ can not be assigned a scaling
dimension. Its correlators decay exponentially and it is
therefore exponentially irrelevant near the UV fixed point.
On the other hand the correlators of the bilinears of the
type $V^*V$ or $V^*\stackrel{\leftrightarrow}\partial_\mu V$
scale with powers of distance. QVD must accomodate
this capricious UV behaviour.

Let us now discuss the fixed point structure of scalar QED.
This is important since if the dual theory is to be exactly
equivalent it must have the same fixed points in its parameter space.

The theory has three couplings: the electromagnetic coupling $e^2$,
the scalar self coupling $\kappa$ and the scalar mass $\mu^2$.
In the following discussion we only consider a physically massless
theory, that is at a given value of $e^2$ and $\kappa$ the mass
$\mu^2$ is tuned in such a way that the theory is critical
(the physical mass vanishes). The remaining two couplings at
every point on a critical surface define two energy scales
$M_\kappa$ and $M_{e^2}$.
Physically at energies much higher than $M_{e^2}$ the
electromagnetic interaction is negligibly small, while far above
$M_\kappa$ the scalar coupling becomes negligible. Naively by
taking each one of these scales to infinity (or rather to the UV
cutoff $\Lambda$ which is infinite from the point of view
of the continuum) or to zero one obtains a theory without a scale.
It is usually the case that such a scale invariant theory is also
conformally invariant and corresponds to a fixed point of
the renormalization group flow. In the present model one therefore
expects four fixed points: A. $M_{e^2}=M_{\kappa}=0$,
B. $M_{e^2}\rightarrow \infty(\propto \Lambda)$,
$M_\kappa\rightarrow \infty(\propto \Lambda)$; C. $M_{e^2}=0$,
$M_\kappa\rightarrow \infty(\propto \Lambda)$;
D. $M_{e^2}\rightarrow \infty(\propto \Lambda)$, $M_\kappa=0$.
With some modifications these are indeed the fixed points of
the Abelian Higgs model.
A. $M_\kappa=M_{e^2}=0$. A theory of noninteracting charges and
free massless photon.  Small changes in both couplings change the
low energy physics and therefore the perturbations associated with
both couplings are relevant. On the two dimensional critical surface
both directions are relevant and this is the UV fixed point.
This of course follows directly from the perturbation theory.
In terms of the bare couplings $e^2=\kappa=0$. It is customary
to characterize the fixed points not in terms of bare
dimensionful couplings but rather in terms of dimensionless
physical couplings. Those are defined as dimensionless Green's
functions at a given scale. For example
\begin {equation}
\kappa_R(\mu)\equiv \mu^7G^4(p_1...p_4)|_{p_ip_j=1/2\mu^2\delta_{ij}}, \  \
e^2_R(\mu)\equiv\mu^7 G^4_{\nu\nu}(p_1...p_4)|_{p_ip_j=1/2\mu^2\delta_{ij}}
\label{renc}
\end{equation}
where $G^4$ and $G^4_{\mu\nu}$ are the connected Green's
functions of four scalar fields and two scalar and two vector
fields respectively. At a fixed point the dependence on the
scale $\mu$ of course disappears. The UV fixed point then
is $e^2_R=\kappa_R=0$.

B.  $M_{e^2}\rightarrow \infty(\propto \Lambda)$,
$M_\kappa\rightarrow \infty(\propto \Lambda)$. On dimensional
grounds it is clear that this fixed point is achieved when both
bare couplings are of the order of UV cutoff $\Lambda$. In terms
of dimensionless couplings $e^2_R=e^2_C$, $\kappa_R=\kappa_C$.
This is the IR fixed point of the theory. The reason is that
once all the couplings of the theory are of the order of the UV cutoff,
the low energy physics should be insensitive to small changes in
any of the couplings
(provided of course, the mass is suitably changed so that the
theory remains massless).
All the directions on the critical surface are thus irrelevant
and the fixed point is IR stable. The existence of this IR fixed
point has been a matter of some controversy. It is now however
well established that the theory does indeed have a second
order phase transition, and therefore an IR fixed point exists.
The evidence comes from lattice simulations \cite{dasgupta},
analytic calculations on the lattice \cite{kleinert}, 1/N
expansion in the continuum strongly coupled model \cite{cpn2},
and perturbative loop expansion \cite{adrian}. Formally taking
the limit $\kappa\rightarrow\infty$ and then $e^2\rightarrow \infty$
one arrives at a $N\rightarrow 1$ limit of a $CP^{N-1}$ theory.
We will refer to this fixed point therefore as
the $CP^0$ model\footnote{Naively it seems that the $CP^0$ model
does not contain any degrees of freedom. However it really means
only that the charged degrees of freedom are irrelevant at the
IR fixed point. This is in full agreement with the universality
argument which in the present case suggests that the IR fixed point
should be the XY model of magnetic vortices.}.

C. $M_{e^2}=0$, $M_\kappa\rightarrow \infty(\propto \Lambda)$. Since
the scalar interaction by itself does not induce vector coupling
this fixed point appears at $e^2=0$, $\kappa\propto \Lambda$
(or $e^2_R=0$, $\kappa_R=\kappa_1$).
 The photon is decoupled from the matter fields.
The dynamics of the matter fields is that of strongly interacting
$U(1)$ invariant $\phi^4$. This just means that the scalar theory
is defined at its IR fixed point, which is the 3D XY model. Clearly,
the direction of the coupling $\kappa$ near this point is irrelevant,
whereas $e^2$ is relevant. So in the space of parameters of QED
this fixed point is stable in one direction which is the $e^2=0$ axis.

 D. Naively one expects to find a fixed point at
$M_{e^2}\rightarrow \infty(\propto \Lambda)$, $M_\kappa=0$.
However in the present model it is impossible to lower the scalar
coupling scale $M_\kappa$ (or equivalently $\kappa_R$) all the way
to zero at nonzero $M_{e^2}$ (or $e^2_R$). The reason is that
for $\kappa_R<\kappa_T$ the vector coupling induces interactions
that change the order of the phase transition from second to
first. The critical surface in the parameter space of QED$_3$
is therefore not infinite but ends on a line of tricritical points.
This tricritical line naturally also ends at a fixed point -
the tricritical fixed point $e^2_R=e^2_T$, $\kappa_R=\kappa_T$.
The existence of this point has been shown analytically \cite{kleinert}
and numerically its measured critical exponents are consistent with
gaussian critical exponents, which is what one expects in 3D
\cite{bartholomew}. This fixed point then must have one irrelevant
direction - the direction along the tricritical line on the
critical surface. The direction corresponding to the flow from
the tricritical point $(e^2_T, \kappa_T)$ to the IR fixed point
$(e^2_C, \kappa_C)$ is relevant.

The phase diagram and the directions of the RG flow near
different fixed points is schematically depicted on Fig.1.
\begin{figure}
%\epsfbox{fig.eps}
\caption{Schematic phase diagram of the Abelian Higgs model.
The arrows show the direction of the renormalization group flow.
The point A is UV stable and has two relevant directions.
The point B is IR stable and has no relevant directions.
The point C corresponds to decoupled photon. The point D is the
tricritical fixed point and the direction along the tricritical
line AD is irrelevant.}
\label{f1}
\end{figure}

\section{A Candidate Theory.}

Consider the following Lagrangian
\begin{equation}
L_{dual}=-\frac{1}{4 g^2}G_{\mu\nu}^2+|(\partial_\mu+iB_\mu)V|^2-
\frac{m^2}{2}B_\mu^2 -M^2V^*V-\lambda(V^*V)^2
\label{duall}
\end{equation}
where $V(x)$ is a scalar complex field, $B_\mu$
a real vector field and $G_{\mu\nu}=\partial_{[\mu} B_{\nu]}$. The field $V(x)$
should be thought of as
a vortex operator of QED$_3$ while $B_\mu$
is closely related to the dual field strength $\tilde F_\mu$.
We shall argue in this section that in a certain strong coupling limit
this theory is equivalent to scalar QED$_3$.

The first thing to note about the Lagrangian eq.(\ref{duall})
is that the mass term of the vector field breaks explicitly the
$U(1)$ "gauge" symmetry. Therefore the field $V$ is a local physical
field and the $U(1)$ symmetry indeed does have a local order
parameter. This is what one expects if this symmetry is
to be identified with the flux symmetry of QED.

As usual a continuum model is defined in the critical region of
the second order phase transition. In the present case it is the
breaking of this $U(1)$ symmetry that is responsible for the
existence of the continuum limit. The fact that the continuum limit
does exist at least for some values of parameters can be
demonstrated in perturbation theory.
For small values of the couplings $g$ and $\lambda$ the
theory is renormalizable perturbatively \cite{collins}.
This can be easily shown as follows.
Let us perform a canonical transformation that decouples the
longitudinal part of the vector field $B_\mu$ \begin{equation}
V=\bar Ve^{-i\frac{\partial_\mu}{\Box}B_\mu}
\label{can}
\end{equation}
The field $\bar V$ now interacts only with the transverse part
of $B_\mu$ and therefore only the transverse part of the
propagator of $B_\mu$ enters the Feynman diagrams of the
theory. But the UV asymptotics of the transverse part of the
massive vector propagator is the same as that of the photon
propagator in QED$_3$ in the Landau gauge. Therefore as far as
the UV asymptotics are concerned, the correlators of $\bar V$ and
transverse $B_\mu$ are the same as in Abelian Higgs model and
are renormalizable. To get back to the correlators of the
original field $V(x$ one just has to undo the canonical
transformation eq.(\ref{can}). Since $\partial_\mu B_\mu$
is a noninteracting field this is straightforward.
{}From the Lagrangian we have
\begin{equation}
<\partial_\mu B_\mu (x) \partial_\nu B_\nu (y)>=
-\frac{\Box}{m^2} \delta^3(x-y)
\end{equation}
Therefore for any $U(1)$ invariant correlator of the field $V$ one has
\begin{eqnarray}
&<V(x_1)...V(x_n)V^*(y_1)...V^*(y_n)>=
<\bar V(x_1)...\bar V(x_n)\bar V^*(y_1)...\bar V^*(y_n)>& \\ \nonumber
&\exp\left\{-\frac{n
\Lambda}{4\pi m^2}+\sum_{i,j}\frac{1}{4\pi m^2|x_i-y_j|}-\sum_{i\ne j}
[\frac{1}{4\pi m^2|x_i-x_j|}+\frac{1}{4\pi m^2|y_i-y_j|}]\right\}&
\label{uvas1}
\end{eqnarray}
where $\Lambda$ is the ultraviolet cutoff.
Evidently multiplying the vortex operator $V$ by the wave
function renormalization factor $Z=\exp\{\frac{\Lambda}{8\pi m^2}\}$
makes all the correlators finite. The theory is therefore
perturbatively renormalizable.

It is however not the weakly coupled limit of eq.(\ref{duall})
that we have in mind as a candidate for QVD.
The reason is that for weak coupling the UV fixed point of
the model is at zero
$g^2$ and $\lambda$ and is a free theory of vortices
rather than a free theory of charges. However the
discussion in a previous section suggests
near which UV fixed point one has to consider eq.(\ref{duall}).
First of all, obviously in the ultraviolet the mass term of
a vector field is unimportant and therefore the UV fixed point
must be among one of the fixed points of scalar QED discussed
earlier. Now, from all the fixed points of QED$_3$ only the
"tricritical" point D is a good candidate. By a universality
argument the tricritical point of QED must be a free theory
of magnetic vortices. This is also consistent with lattice
simulations \cite{bartholomew}. But after omission of the mass term,
the Lagrangian eq.(\ref{duall}) describes a gauge interaction of
vortices. So in the Lagrangian eq.(\ref{duall}) the roles
of charges and vortices  are interchanged relative to the
original Lagrangian of scalar QED$_3$. Therefore the UV fixed point
relevant to our discussion must be the point D with the
obvious change of $e^2$ into $g^2$ and $\kappa$ into  $\lambda$.
It describes free electric charges. QVD should therefore be
defined by eq.(\ref{duall}) with couplings near the fixed point
$g^2_R=e^2_T$ , $\lambda_R=\kappa_T$ (the renormalized couplings $g_R^2$ and
$\lambda_R$ are defined by eq.(\ref{renc}) interchanging the charged field into
the vortex field and the vector potential into the vector field $B_\mu$). In
terms of the bare
couplings it is a formal limit $g^2\rightarrow \infty$,
$\lambda\rightarrow\infty$ with the ratio $g^2/\lambda$ finite and fixed.

The perturbative renormalizability tells us that the theory
remains renormalizable also at strong bare couplings as long
as both theories are connected by a critical line. The physical
picture is simple. The renormalized theory with small couplings
$g$ and $\lambda$ has a certain energy scale $M_g$ associated with
the vector  coupling. Above this scale the asymptotic UV behaviour
sets in and the correlators are perturbatively calculable. Below this scale
the perturbation theory breaks down for small enough physical mass since
the effective dimensionless expansion parameter is $g^2/M$. However
the renormalizability of the theory assures that all the correlators are
finite also for these momenta. In perturbation theory this
crossover scale is of order $M_g\propto g^2$. The limit that
interests us involves raising this crossover scale to the scale
of the UV cutoff and simultaneously fine tuning $\lambda$
so that the effective quartic scalar self coupling at $M_g$ be
as small as possible. However since this is done already on
the level of renormalized theory, the correlators remain
finite in this limit. It is therefore reasonable to expect
that going from the weakly coupled massive scalar QED to
QVD will not introduce new UV divergencies
and the theory will remain UV finite.

Regarding the second condition of the previous section, it is
evidently satisfied by the Lagrangian eq.(\ref{duall}). The kinetic
term of the vector field $B_\mu$ is negligible at low energies
and the field $B_\mu$ for the purposes of low energy physics can
be expressed in terms of the vortex field $V(x)$
\begin{equation}
B_\mu=\frac{i}{-m^2+2V^*V}V^*\stackrel\leftrightarrow\partial_\mu V
\end{equation}
Substituting it back in the Lagrangian one finds
\begin{equation}
L_{dual}=\partial_\mu V^*\partial_\nu V +\frac{1}{-2m^2+4V^*V}
(V^*\stackrel\leftrightarrow\partial_\mu V)^2-M^2V^*V-\lambda(V^*V)^2
\label{vecvec}
\end{equation}
In the infrared region the interaction terms which involve
derivatives become less important than the potential terms and
therefore the theory degenerates into the Landau theory eq.(\ref{landau}).

Now let us discuss the limit of QVD which reproduces main features
of the $e^2\rightarrow 0$ limit in QED$_3$. This will also help
us to establish roughly the correspondence between some of the
couplings in eq.(\ref{duall}) with the QED couplings. To this end
let us calculate the infrared divergent contribution to the
energy of a hedgehog soliton in the phase where $<V>=v\ne 0$.
Asymptotically the hedgehog configuration is
\begin{equation}
V(x)=ve^{i\theta}, B_i=-\frac{2v^2}{-m^2+2v^2}\partial_i\theta(x)
\label{sol}
\end{equation}
where $\theta(x)$ is a planar angle function. Assuming
$m^2<<v^2$ (that is that
we are far from the phase transition line, which is
the region where the perturbation theory in QED is valid), we obtain
\begin{equation}
E_{soliton}=\frac{\pi m^2}{2} \ln (LM)+{\rm IR \ finite}
\end{equation}
where $L$ is the infrared cutoff. The hedgehog soliton in the
dual theory represents the charged particle of QED$_3$. But for
weak coupling the energy of the charged scalar in QED$_3$ also
diverges in the IR as $\frac{e^2}{8\pi}\ln{L}$. We find therefore
that to lowest order in $e^2$ the following relation
between the couplings should hold
\begin{equation}
m^2=\frac{e^2}{4\pi^2}
\label{cor}
\end{equation}
The $e^2\rightarrow 0$ limit in QED$_3$ therefore should
be likened to the
$m^2\rightarrow 0$ limit in QVD.

Indeed the $m^2\rightarrow 0$ limit is singular. In this limit the $U(1)$
phase rotation in the dual theory eq.(\ref{duall}) becomes
a local gauge symmetry. The field $V(x)$ becomes
then nongauge invariant.
Moreover, since we are interested in the limit
$g^2\propto\Lambda$, the gauge field $B_\mu$ does not have a
kinetic term at any finite momentum  and the global $U(1)$
transformations do not generate a global symmetry but rather a
part of the gauge group and act trivially on all physical
states \cite{cpn}. The model still has however a conserved
topological current
$J_\mu=\epsilon_{\mu\nu\lambda}\partial_\nu B_\lambda$.
A solitonic configuration  that carries a unit of this topological
charge eq.(\ref{sol}) at $m^2=0$ differs from
vacuum  asymptotically only by a gauge transformation. Therefore the soliton is
now a local field configuration and the operator creating
the soliton becomes a local operator.
The soliton energy accordingly becomes IR finite.
As for the elementary excitations of eq.(\ref{duall}), the vortices are not
interpolated by $V$. To interpolate them one would
have to construct some gauge invariant operator with the same global
quantum numbers. However as discussed in \cite{cpn} such an operator
does not exist precisely due to the fact that the gauge field does not
have a kinetic term. There still are excitations that carry the vorticity
$\epsilon_{\mu\nu\lambda}\partial_\nu J_\lambda$ \cite{cpn}.
There is a "transmutation" of quantum numbers. The elementary
excitations instead of carrying the global U(1) magnetic flux
carry only vorticity. They are however nonlocal configurations of
gauge invariant fields and are logarithmically confined.
This picture is precisely
 the same as described in item 3 of the previous section.

The smallness of $m^2$ for small $e^2$ tells us also another thing.
Let us concentrate on a phase in which $<V>=0$. In that case the
coefficient of the vector-vector interaction in eq.(\ref{vecvec}) is
proportional to $1/e^2$ and is very large. The vortex and antivortex
therefore attract each other strongly. Therefore the formation of
tightly bound vortex-antivortex bound states is favored by the
dynamics. Since the $<V>=0$ phase in this model is dual to the Higgs
phase of scalar QED$_3$ these bound states represent the massive
photon and the massive Higgs particle.

The last point concerns the UV behavior of different correlators.
As mentioned in the previous section, the correlators of
vortex operators in QED  have exponential rather than power
UV asymptotics, see eq.(\ref{uvas}). But as shown in the discussion
of the renormalization in the dual theory this is precisely the UV
asymptotic behavior of the correlators of $V(x)$ in eq.(\ref{duall}).
In fact even the coefficient in the exponential in eq.(\ref{uvas}) is
the same as in eq.(\ref{uvas1}) if the correspondense between
$e^2$ and $m^2$ of eq.(\ref{cor}) is used.
Moreover, the composite operators of the type $V^*V$ are
invariant under the canonical transformation,
eq.(\ref{can}) and therefore scale in the same way as
$\bar V^* \bar V$, that is have a power law behavior.
The operator $V^*\partial_\mu V$ under the canonical transformation
eq.(\ref{can}) transforms into $\bar V^*\partial_\mu
\bar V-i\bar V^*\bar V\frac{\partial_\mu\partial_\nu}{\Box}B_\mu$.
However both, polynomials of $\bar V^*\bar V$ and  of $B_\mu$
scale as powers therefore
this operator also has a power law behavior in the UV.
Evidently this argument holds for any operator which is
invariant under global $U(1)$ transformation, since under
the canonical transformation, eq.(\ref{can}), it will transform
into a finite polynomial of $\bar V$, $B_\mu$ and their derivatives.
We then conclude that any local operator invariant under global $U(1)$
transformation in the dual model, eq.(\ref{duall}), scales in the
UV region as a power of distance, but any eigenoperator of $U(1)$
with nonzero eigenvalue scales as an exponential times a power. This
is precisely the behavior expected from the perturbation theory in QED$_3$.

The dual theory we discussed is strongly coupled and therefore is
not easily accessible to a quantitative analysis. One can give
however several qualitative arguments which elucidate further
the connection between this model and the Higgs model and support
their conjectured equivalence. Based on very general assumptions
we discuss now the fixed point structure of QVD, eq.(\ref{duall}), and
argue that there is a one to one correspondence between its fixed
points and the fixed points of QED$_3$ and that this
correspondence is totally consistent with duality.

i. The UV fixed point. As we have discussed earlier QVD is
defined at the UV fixed point $g^2_R=e^2_T$, $\lambda_R=\kappa_T$.
Being dual to the tricritical point of QED$_3$ the theory must
therefore descibe free charges. However if this is to be
equivalent to UV point of QED$_3$ it must also contain a
noninteracting massles pseudoscalar particle - the QED photon.
In fact it is easy to see that this is correct. QED$_3$
is defined initially with the zero mass of the
vector field and the longitudinal part of the photon
(which at zero mass is completely decoupled) is thrown out
of the Hilbert space. The UV fixed point of QVD on the other
hand is the limit $m^2\rightarrow 0$ of eq.(\ref{duall}).
At any finite $m$ the longitudinal part of the vector field
$B_\mu$ is in fact in the Hilbert space of the theory. In the limit
$m^2\rightarrow 0$ it is decoupled but still exists. Since $B_\mu$
represents the dual field strength of QED, it is a pseudovector
field. Its longitudinal part for $m=0$ describes therefore
a decoupled massless pseudoscalar particle which
evidently can be thought of as a photon.

ii. The IR fixed point. In the IR limit all dimensional
couplings are of the order of the UV cutoff. Therefore we
have $g^2\propto\Lambda$, $\lambda\propto\Lambda$,  $m^2\propto\Lambda$.
The mass of a vector particle behaves as $m_V^2\propto g^2m^2$.
The renormalized vector coupling then scales as
$g^2_R(\mu)\propto g^2\mu^3/m_V^4=\mu^3/g^2m^4\rightarrow 0$.
The vector field therefore completely decouples from the
scalar and also becomes infinitely heavy.
The theory then becomes just the strongly interacting $(V^*V)^2$
model - the XY model for the vortex field. As we have
discussed in the previous section this is precisely what one
expects as the IR fixed point of
scalar QED$_3$ on the grounds of universality.

iii. Taking the limit $m^2\rightarrow 0$ and also
$\lambda_R=\kappa_C$, $g^2_R=e^2_C$, QVD moves to a
fixed point dual to the point C of QED. Now it is the CP$^0$
model of the vortex field. Additionally this fixed point
contains a massless pseudoscalar particle - the longitudinal
component of $B_\mu$. If indeed the CP$^0$ model of charges
is equivalent to the XY model of vortices, then
by duality this
fixed point must be equivalent to the XY model of charges plus
a decoupled photon.

iv. Finally one can take $m^2$ to infinity at a fixed $g^2$.
This again leads to $g^2_R=0$. Then the vector field disappears
from the spectrum and decouples from the scalar. It is then possible
to take the limit $\lambda\rightarrow 0$, which gives a
theory of free noninteracting massless vortices. This is the
tricritical point and should be equivalent to the point D in the figure.

The correspondence between the fixed points of QED and QVD
is summarized in Table 1.
\begin{table}
\begin{center}
\begin{tabular}{||c||c||c||}
\hline \hline
Fixed point & QED$_3$  & QVD$_3$ \\ \hline  \hline
UV & Free charges + photon, & Strongly interacting vortices\\
  & $e^2_R=0$, $\kappa_R=0$  & $g^2_R=e^2_T$, $\lambda_R=\kappa_T$,
$m^2=0$\\ \hline
IR & Charge CP$^0$, & Vortex XY\\
&  $e^2_R=e^2_C$, $\kappa_R=\kappa_C$ &$g^2_R=0$,
$m^2\rightarrow\infty$, $\lambda_R=\kappa_1$ \\ \hline
Tricritical & Strongly interacting charges & Free vortices\\
&$e^2_R=e^2_T$, $\kappa_R=\kappa_T$ & $g^2_R=0$,
$m^2\rightarrow\infty$, $\lambda=0$ \\ \hline
End point & Charge XY + decoupled photon, & Vortex
CP$^0$ + decoupled photon \\
& $e^2_R=0$, $\kappa_R=\kappa_1$ &$g^2_R=e^2_C$,
$\lambda_R=\kappa_C$, $m^2=0$ \\ \hline  \hline
\end{tabular}
\end{center}
\label{table}
\caption{Fixed points of Abelian Higgs model in terms of
QED$_3$ and QVD$_3$ couplings}
\end{table}

\section{Discussion.}

In this note we have presented a continuum field
theory which has many features
similar to scalar QED$_3$. Those include the phase
structure of the theory, the
infrared asymptotics, the leading ultraviolet behavior of
correlators and the fixed point structure and the renormalization group
flow on the critical surface. On the basis of
these similarities we conjecture that this theory
which we call QVD is equivalent to
scalar QED$_3$ on all energy scales.
It should be noted that the QVD Lagrangian
is written explicitly in terms of
local gauge invariant fields. The gauge dynamics in
2+1 dimensions is thereby described without the use
of nongauge invariant vector potentials.

Since the number of relevant directions at all fixed points
of QVD and QED$_3$ is the same there should be a one to one
mapping between the parameters of QVD and QED which maps
one theory exactly into the other.
It would be very interesting to find this mapping explicitly.
Unfortunately in the region of parameter space in which QED
is weakly interacting, the interactions of the vortex field
in QVD is strong and vice versa. Finding this exact mapping
is therefore a very involved matter and we do not have adequate
analytic tools to do so. Several qualitative features can
however be determined. We have shown for example that for weak
gauge couling the vector mass term in QVD is proportional to
the electromagnetic coupling $m^2=e^2/4\pi^2$. One can also
calculate the mass of the magnetic vortex in the framework of
QED semiclassically in the Higgs phase far from the phase
transition \cite{preskill}. One then gets the following
relation between the scalar
mass coefficient of QVD and the QED couplings:
\begin{equation}
M=\frac{\pi\mu^2}{4\kappa}{\rm ln} (e^2/\kappa)
\end{equation}
Reversing the argument and performing the analogous
calculation deep in the Coulomb phase in the limit
$m^2\rightarrow 0$ in QVD
one obtains the mass of the charged particle as
\begin{equation}
\mu=\frac{\pi M^2}{4\lambda}{\rm ln} (g^2/\lambda)
\end{equation}

Unfortunately since the semiclassical approximation is
only valid in a small region of parameter space and since we
are dealing with the strongly interacting theories the bulk of
our discussion has been qualitative. It would be very
interesting to check numerically the equivalence picture
presented here. Although of course the numerical proof of
equivalence is impossible, it is nowadays possible to measure
scaling dimensions of various operators in computer simulations
with great accuracy. It would be ineresting therefore to
verify the main features of the RG flow and scaling laws
discussed here in a large scale computer simulation.

{ \large \bf Acknowledgements}
We thank H. Kleinert, A. Krasnitz and A.M.J. Schakel
for interesting discussions.

\end{document}